\documentclass[12p]{article}
\usepackage{latexsym}
\usepackage{color}
\usepackage{amsmath}
\usepackage{epsfig}
\usepackage{hyperref}
 \usepackage{subfigure}
 \usepackage{dcolumn}
   \usepackage{threeparttable}

\newcommand{\ortala}[1]{\begin{center}#1\end{center}}

\newcommand{\ket}[1]{\left|#1\right\rangle}

\newcommand{\sand}[3]{\left\langle #1\left|#2\right|#3\right\rangle}
\newcommand{\sandd}[1]{\left\langle #1\right\rangle}

\newcommand{\summ}[3]{{{\underset{#1 }{\overset{#2}{\displaystyle\sum}}}#3}}
\newcommand{\prodd}[3]{{{\underset{#1
}{\overset{#2}{\displaystyle\prod}}}#3}}

\newcommand{\re}[1]{(\ref{#1})}

\newcommand{\eq}[2]{\begin{equation}\label{#1}  #2\end{equation}}

\newcommand{\paran}[1]{\left(#1\right)}

\newcommand{\sch}[1]{Schrodinger}
\newcommand{\ktur}[2]{\frac{\partial #1}{\partial #2}}

\newcommand{\komb}[2]{\paran{\begin{array}{c} #1 \\ #2 \end{array}}}

\usepackage{amssymb}
\usepackage{latexsym}
\begin{document}

\ortala{\large\textbf{Hysteresis behaviors of a spin-1 anisotropic Heisenberg model}}

\ortala{\textbf{\"Umit Ak\i nc\i \footnote{\textbf{umit.akinci@deu.edu.tr}}}}

\ortala{\textit{Department of Physics, Dokuz Eyl\"ul University,
TR-35160 Izmir, Turkey}}

\section{Abstract}

The hysteresis behaviors of anisotropic S-1 Heisenberg model have been studied within the effective field theory with two spin cluster. After giving the phase diagrams, the effect of the crystal field and anisotropy in the exchange interaction on the hysteresis loops have been determined. One important finding is double hysteresis loop behavior of the system in the low temperature and negative crystal field region, which disappears with the decreasing anisotropy in the exchange interaction. This behavior was carefully investigated and physical explanation has also been given.

\section{Introduction}\label{introduction}

Spin-1 (S-1) Heisenberg model has attracted some interest for many years. The model has exchange and single-ion anisotropies. There are many materials which can be handled by this type of model e.g. $K_2NiF_4$ \cite{ref_1}, $CsNiCl_3$ \cite{ref_2}, $CsFeBr_3$ \cite{ref_3}, which have weak axial anisotropy and strong planar anisotropy, respectively. Besides, molecular oxygen $O_2$ adsorbed on graphite \cite{ref_4},  $NiGa_2S_4$ \cite{ref_5}, and $Ba_3Mn_2O_8$ \cite{ref_6} are the more recent examples of this type of materials. Indeed, in real magnetic materials, single-ion anisotropy plays a major role for investigating the magnetic behavior of the system \cite{ref_7}. On the other hand, it has been shown from the band structure calculations that, spatially anisotropic exchange interaction along the x and y directions occur in several  vanadium phosphate material systems, such as $(Pb_2VO(PO_4))_2$, $SrZnVO(PO_4)$, $BaZnVO(PO_4)$ and $BaCdVO(PO_4)_2$ \cite{ref_8}.

S-1 anisotropic or isotropic Heisenberg model has been studied by various methods, in order to obtain phase transition characteristics and thermodynamical properties. We can inspect the related literature by grouping the works as ground state works and finite temperature works.

The ground-state properties of this system has been worked widely and quantum phase transition characteristics have been obtained.
For instance, S-1 Heisenberg ferromagnet with an arbitrary crystal-field potential within the linked-cluster series expansion \cite{ref_9}, S-1 Heisenberg antiferromagnet with uniaxial single-ion anisotropy in a field within the spin-wave approach \cite{ref_10}, two-dimensional quantum anisotropic S-1 Heisenberg antiferromagnet using the self consistent harmonic approximation \cite{ref_11}, S-1 Heisenberg antiferromagnetic spin chains with exchange and single-site anisotropies in an external field  within the density-matrix renormalization group techniques \cite{ref_12},  S-1 bilinear-biquadratic model on a honeycomb lattice with using the linear flavor-wave (LFW)
theory \cite{ref_13} and   tensor renormalization group method \cite{ref_14}, S-1 bond-alternating Heisenberg
antiferromagnetic chain with a single-ion anisotropy in longitudinal and transverse magnetic fields within  the  infinite time evolving block decimation, the linearized tensor renormalization group, and
the density matrix renormalization group methods \cite{ref_15},
S-1 antiferromagnetic Heisenberg model on the square lattice with a three-site interaction using a variety of analytical and numerical methods \cite{ref_16} and S-1  Heisenberg model with a single-ion  anisotropy on a triangular lattice within the
the cluster mean-field approach \cite{ref_17}.

With all these works, the nature of the ground state of the S-1 Heisenberg model has been determined.  Recently it has been shown that, the quantum effects can also be seen in a finite (but low enough) temperature region \cite{ref_18}. Hence,  the literature also include  the works related to the quantum phase transitions in the region of low temperatures, in which the quantum effects are certainly still present. For example, the effect of the magnetic field on the phase diagrams and the thermodynamical properties has been studied
for a S-1 bilinear-biquadratic Heisenberg model on the triangular lattice within the mean-field theory and exact diagonalizations \cite{ref_19}, S-1 Heisenberg antiferromagnet with easy-axis or easy-plane single-site anisotropy on the square lattice within the
series expansions \cite{ref_20},
the effects of frustration between nearest, next-nearest neighbor and next-next-nearest
neighbors of the quantum  S-1 anisotropic antiferromagnetic Heisenberg model on a simple
cubic lattice with single ion anisotropy using the bond operator technique \cite{ref_22}, are among them. Most recent studies are, the phase diagram of the anisotropic S-1 Heisenberg chain with single ion anisotropy (D) using a ground-state fidelity approach \cite{ref_23}, the low-temperature properties of one-dimensional S-1 Heisenberg model with geometric
fluctuations  within the strong-disorder renormalization-group
and  quantum Monte Carlo and density-matrix renormalization-group numerical calculations \cite{ref_24} and two-dimensional S-1 antiferromagnet with next nearest neighbor exchange interactions and easy axis single ion anisotropy, on the square lattice, are studied
at low temperature using a Modified Spin Wave Theory \cite{ref_25}.

On the other hand, when we look at the phase diagrams and thermodynamical properties of the system in the finite temperature region, we can see several works with various methods, as in the ground properties of the system.

One of the first studies in this case is,  S-1 uniaxial ferromagnetic model with both exchange anisotropy and single-ion
anisotropy ($D$), as well as transverse coupling are studied in the mean-field approximation \cite{ref_26}. The phase diagrams of the system have been obtained in this work. One-dimensional S-1 ferromagnetic isotropic
Heisenberg model is studied by the double-time Green's function method \cite{ref_27} and two-dimensional classical isotropic Heisenberg antiferromagnet with a single-ion anisotropy is studied in the presence of a uniform
magnetic field along the easy axis within the MC simulation and Green's function technique \cite{ref_28}.

The three components of the magnetizations for the
anisotropic Heisenberg model with single ion anisotropy have been calculated by the use of many-body
Green's function method, when external field is applied in both x and z directions \cite{ref_29}. The effect of an easy-plane crystalline anisotropy and easy-axis exchange anisotropy on the phase diagram of the three-dimensional classical ferromagnetic anisotropic Heisenberg model has been studied by using Monte Carlo simulations \cite{ref_30}. In addition, the thermodynamic properties of S-1 ferromagnetic chains with an easy-axis single-ion anisotropy have been
investigated by both a Green-function approach, based on a decoupling of three-spin
operator products, and by exact diagonalizations of chains with up to $N = 12$ sites using periodic boundary
conditions \cite{ref_31}.

The general works related to spin S Heisenberg systems have also been present.
The thermodynamical properties  of one and two-dimensional ferromagnets with arbitrary spin-S in a magnetic
field have been investigated by a second-order Green-function theory for the isotropical Heisenberg model \cite{ref_32}.
Also  the phase diagrams of the anisotropic ferromagnetic spin-S Heisenberg model on a square lattice have been obtained by double-time Green's function method within the Callen decoupling approximation \cite{ref_33}.

Most recent studies related to the thermodynamic properties and phase transition characteristics are,  S-1 Heisenberg antiferromagnet with easy-plane single-ion anisotropy on three-dimensional bipartite lattices with sixth-order series expansions \cite{ref_34}, three-dimensional anisotropic Heisenberg XXZ model with a crystal field by using the variational approach based on the Bogoliubov inequality \cite{ref_35} and quantum S-1 anisotropic ferromagnetic Heisenberg model within the same method \cite{ref_36}.
The field-induced laws of thermodynamic properties are obtained by Green's function method for the two-dimensional S-1 ferromagnetic Heisenberg model with the exchange and single-ion anisotropies \cite{ref_37}. Some exact solutions were aslo reported such as  S-1 Ising-Heisenberg diamond chain in a magnetic field by a rigorous treatment based on the transfer-matrix method \cite{ref_38}.

As seen in this short literature, the problem of thermodynamical properties of the S-1 anisotropic Heisenberg model is up to date. Thus,
the aim of this work is to determine the hysteresis behaviors of the S-1 anisotropic Heisenberg model on a bulk system. For this aim, the paper is organized as follows: In Sec. \ref{formulation} we
briefly present the model and  formulation. The results and
discussions are presented in Sec. \ref{results}, and finally Sec.
\ref{conclusion} contains our conclusions.

\section{Model and Formulation}\label{formulation}

We start with a standard S-1 anisotropic Heisenberg Hamiltonian as,
\eq{denk1}{
\mathcal{H}=-J\summ{<i,j>}{}{\left[\Delta\paran{ S_i^xS_j^x+ S_i^yS_j^y}+ S_i^zS_j^z\right]}-D\summ{i}{}{\paran{S_i^z}^2}-H\summ{i}{}{S_i^z}
}
where $S_i^x,S_i^y,S_i^z$ denote the $x,y,z$ component of the Pauli S-1 operator at a site $i$ respectively,  $J$ stands for the exchange interactions between the nearest neighbor spins, $\Delta$ is the anisotropy in the exchange interaction, $D$ and $H$ are the crystal field and longitudinal external magnetic field  at sites of the lattice, respectively. The first summation is carried over the nearest neighbors of the lattice, while the others are over all the lattice sites. We note that, Hamiltonian given in Eq. \re{denk1} covers Ising model, anisotropic Heisenberg model and isotropic Heisenberg model depending on the value of $\Delta$. When $\Delta=0$, Eq. \re{denk1} represents the S-1 Ising model (Blume-Capel model) while $\Delta=1$ gives the isotropic S-1 Heisenberg model. The intermediate values of the anisotropy in the exchange interaction ($0<\Delta<1$) corresponds to the anisotropic S-1 Heisenberg model, by means of the $XXZ$ model.

In an  EFT-$2$ approximation \cite{ref_39}, we start by constructing the $2$-site cluster and write $2$-site cluster Hamiltonian with the axial approximation \cite{ref_40} as
\eq{denk2}{
\mathcal{H}^{(2)}=-J\left[\Delta\paran{ S_1^xS_2^x+ S_1^yS_2^y}+ S_1^zS_2^z\right]-
D\summ{i=1}{2}{\paran{S_i^z}^2}-\summ{i=1}{2}{\paran{h_i+H}S_i^z}
}
Here $h_i$  denotes all the interactions between the  spin at a site $i$ and the other spins which are outside of the cluster, where $i=1,2$.   Let the site $i$ has number of $z_i$ nearest neighbor sites which are located at the outside of the selected cluster, then $h_i$ can be written  within the axial approximation \cite{ref_40} as
\eq{denk3}{
h_i=J\summ{k=1}{z_i}{S_{i,k}^z},
} where $S_{i,k}^z$ is the $k.$ the nearest neighbor of the spin $S_i^z$.
The thermal average of  the quantity $S_i^z$ ($i=1,2$) via exact generalized
Callen-Suzuki identity \cite{ref_41} is given by
\eq{denk4}{
\sandd{S_i^z}=\sandd{\frac{Tr_2 S_i^z \exp{\paran{-\beta \mathcal{H}^{(2)}}}}{Tr_2 \exp{\paran{-\beta \mathcal{H}^{(2)}}}}}
.} In Eq.  \re{denk4},  $Tr_2$ stands for the partial trace over the lattice sites $1$ and $2$,  which belong to the constructed finite cluster and $\beta=1/(k_BT)$, where $k_B$ is the Boltzmann constant and $T$ is the temperature. Calculation of Eq. \re{denk4} requires the matrix representation of the related operators in chosen basis set, which can be denoted by $\{\ket{\phi_i\}}$ where $i=1,2,\ldots, 9$. Each of the element of this basis set can be represented by   $\ket{s_1s_2}$, where $s_i=\pm 1,0 $ is just one spin eigenvalues of the S-1 operator $S_i^z$  ($i=1,2$).

Let us denote the  matrix elements of the operator $\mathcal{H}^{(2)}$ defined in Eq. \re{denk2} by,
\eq{denk5}{
R_{ij}=\sand{\phi_i}{\mathcal{H}^{(2)}}{\phi_j}.
}These elements can be found in Sec. \ref{app_a}.

In order to get the matrix representation of the operator $\exp{\paran{-\beta \mathcal{H}^{(2)}}}$,
the matrix whose elements defined in Eq. \re{denk5} has to be diagonalized. But the diagonal form of this matrix cannot be obtained analytically, thus some numerical procedures have to be applied. After numerical diagonalization, diagonal elements of this matrix  $r_\alpha$ (where $\alpha=1,2,\ldots,9$) can be obtained. Needless to say, these are just the eigenvalues of the matrix whose elements are defined in Eq. \re{denk5}. Let us denote the transformed basis set  as $\ket{\phi_i^\prime}$  ($i=1,2,\ldots,9$) which makes the matrix representation of the operator $\mathcal{H}^{(2)}$ diagonal,

\eq{denk6}{
r_{\alpha}=\sand{\phi_\alpha^\prime}{\mathcal{H}^{(2)}}{\phi_\alpha^\prime}
.}

Let us denote the diagonal matrix elements of the operator $S_i^z$ in this new basis set as

\eq{denk7}{
T_{i,\alpha}=\sand{\phi_\alpha^\prime}{S_i^z}{\phi_\alpha^\prime}
}

With these definitions, Eq. \re{denk4} can be written as
\eq{denk8}{
\sandd{S_i^z}=\sandd{\frac{\summ{\alpha=1}{9}{}T_{i,\alpha} \exp{\paran{-\beta r_\alpha}}}{\summ{\alpha=1}{9}{}\exp{\paran{-\beta r_\alpha}}}}
.}

The order parameter of the system can be defined by
\eq{denk9}{
m=\frac{1}{2}\paran{\sandd{S_1^z}+\sandd{S_2^z}}
.}

By writing Eq. \re{denk8} in Eq. \re{denk9} we can get the magnetization of the system in a closed form as
\eq{denk10}{
m=\sandd{f\paran{h_1,h_2}}
,}  where the function is defined as

\eq{denk11}{
f\paran{h_1,h_2}=\frac{1}{2}\frac{\summ{\alpha=1}{9}{}\paran{T_{1,\alpha}+T_{2,\alpha}} \exp{\paran{-\beta r_\alpha}}}{\summ{\alpha=1}{9}{}\exp{\paran{-\beta r_\alpha}}}
.}
We note that the function in Eq. \re{denk11}
also the functions of the Hamiltonian variables ($J,\Delta,H,D$) as well as the temperature. Needless to remind that, the function in Eq. \re{denk11} cannot given analytically.

Evaluating Eq. \re{denk10} with using differential operator technique and decoupling approximation \cite{ref_42} is possible. This approximation is most widely used for these type of systems within the effective field theory. Within this technique  Eq. \re{denk10} can be written as

\eq{denk12}{
m=\sandd{\exp{\paran{h_1\nabla_1+h_2\nabla_2}}}f(x_1,x_2)|_{x_1=0,x_2=0}
} where $\nabla_1=\ktur{}{x_1}, \nabla_2=\ktur{}{x_2}$ are the differential operators, and the effect of the exponential differential operator on an arbitrary function $f(x_1,x_2)$ is defined by
\eq{denk13}{
\exp{\paran{a_1\nabla_1+a_2\nabla_2}}f(x_1,x_2)=f(x_1+a_1,x_2+a_3),
} where $a_1,a_2$ are the arbitrary constants.

With writing Eq. \re{denk3} in Eq. \re{denk12} and using S-1 Van der Waerden identity \cite{ref_43}, we can arrive the expression of the order parameter as

\eq{denk14}{
m=\sandd{\prodd{i=1}{2}{}\prodd{k=1}{z_i}{\left[1+S_{i,k}\sinh{\paran{J\nabla_i}+S_{i,k}^2\paran{\cosh{\paran{J\nabla_i}}-1}}\right]}}f(x_1,x_2)|_{x_1=0,x_2=0}
}

The quadrupolar moment of the system which is defined by
\eq{denk15}{
q=\frac{1}{2}\paran{\sandd{\paran{S_1^z}^2}+\sandd{\paran{S_2^z}^2}}=\sandd{\exp{\paran{h_1\nabla_1+h_2\nabla_2}}}g(x_1,x_2)|_{x_1=0,x_2=0}
}
can be calculated in the same way of the magnetization of the system as
\eq{denk16}{
q=\sandd{\prodd{i=1}{2}{}\prodd{k=1}{z_i}{\left[1+S_{i,k}\sinh{\paran{J\nabla_i}+S_{i,k}^2\paran{\cosh{\paran{J\nabla_i}}-1}}\right]}}g(x_1,x_2)|_{x_1=0,x_2=0}
}

where the function defined by
\eq{denk17}{
g\paran{h_1,h_2}=\frac{1}{2}\frac{\summ{\alpha=1}{9}{}\paran{U_{1,\alpha}+U_{2,\alpha}} \exp{\paran{-\beta r_\alpha}}}{\summ{\alpha=1}{9}{}\exp{\paran{-\beta r_\alpha}}}
.} Here $U_{i,\alpha}$ is given by

\eq{denk18}{
U_{i,\alpha}=\sand{\phi_\alpha^\prime}{\paran{S_i^z}^2}{\phi_\alpha^\prime}
.}
We note that by writing Eqs. \re{denk14} and \re{denk16}, we have assumed that the spins $S_1$ and $S_2$ have no common nearest neighbors.

If we use the decoupling approximation \cite{ref_42} in Eqs. \re{denk14} and \re{denk16} and  assume the homogeneity of the lattice (i.e. all lattice sites are equivalent) we get the expressions

\eq{denk19}{
\begin{array}{lcl}
m&=&\Theta_1^{z_1}\Theta_2^{z_2}f(x_1,x_2)|_{x_1=0,x_2=0}\\
q&=&\Theta_1^{z_1}\Theta_2^{z_2}g(x_1,x_2)|_{x_1=0,x_2=0}
\end{array}}

where the operator is defined by
\eq{denk20}{
\Theta_i=\left[1+m\sinh{\paran{J\nabla_i}+q\paran{\cosh{\paran{J\nabla_i}}-1}}\right]
.} Eq. \re{denk19} is a coupled nonlinear equation system of $m,q$. Before the numerical solution, last step contains writing the operators in Eq. \re{denk19} as exponentials in order to use Eq. \re{denk13}. Hence, Eq. \re{denk20} can be written as

\eq{denk21}{
\Theta_i^{z_i}=\summ{n=0}{z_i}{}\summ{p=0}{n}{}C_{np}^{(i)}m^pq^{n-p}
}
with Binomial expansion. Here the coefficients are defined by
$$
C_{np}^{(i)}=\komb{z_1}{n}\komb{n}{p}\summ{r=0}{n-p}{}\summ{t=0}{r}{}\summ{v=0}{p}{}
\komb{n-p}{r}\komb{r}{t}\komb{p}{v}\times
$$
\eq{denk22}{
\paran{\frac{1}{2}}^{r+p}\paran{-1}^{n-p-r+v}
\exp{\left[\paran{r+p-2t-2v}J \nabla_i\right]}.
}

The critical temperature of the system can be obtained by linearizing Eq. \re{denk19} in m.

\newpage
\section{Results and Discussion}\label{results}

In this section, after the phase diagrams of the system is reviewed, hysteresis characteristics of the system will be obtained, for a square lattice ($z_1=z_2=3$) as an example of the 2D lattices. We use the scaled quantities given as
\eq{denk23}{
t=\frac{k_BT}{J}, \quad h=\frac{H}{J}, \quad d=\frac{D}{J}}
within the calculations.

\subsection{Phase diagrams}

\begin{figure}[h]\begin{center}
\epsfig{file=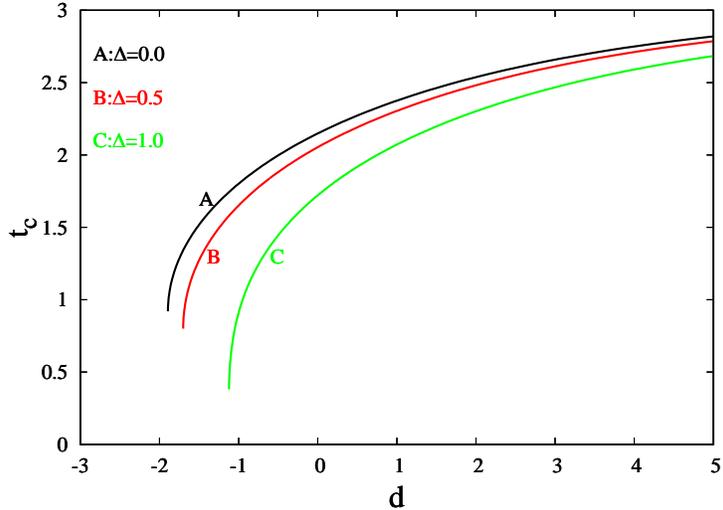, width=10cm}
\end{center}
\caption{The variation of the critical temperature with crystal field for selected values of $\Delta$, for anisotropic S-1 Heisenberg model on a square lattice.} \label{sek1}\end{figure}

The phase diagrams of the square lattice within the S-1 anisotropic Heisenberg model can be seen in Fig. \ref{sek1}, for selected values of anisotropy in the exchange interaction ($\Delta$). We can see from Fig. \ref{sek1} that, all curves related to the different anisotropy in the exchange interactions have tricritical points, at which plotted phase diagrams end. At this point, the second order magnetic phase transition meets the first order one. We note that, when the anisotropy in the exchange interaction decreases (i.e. $\Delta$ increases), the tricritical point shifts towards the low temperature and crystal field region of the $(t_c,d)$ plane. We can say that all curves have the same characteristics, regardless of the value of $\Delta$ : when the crystal field decreases from the positive region, the critical temperature of the system decreases. This decreasing behavior of the critical temperature of the system becomes more rapid in the negative crystal field region. At a specific value of the crystal field, second order critical curves terminated (at specific tricritical values). The isotropical model has lower critical temperature than the anisotropical model, for all values of $d$. Since rising anisotropy in the exchange interaction (decreasing $\Delta$) forces the spins  align along the $z$ axis, then it is not surprising that, the required energy supplied by the temperature to destroy the order of the spins $z$ direction, has to be higher.

We can compare the results in the critical temperatures with other methods for some known limits of the model such as $d\rightarrow \infty$ (namely Ising limit), $d=0$. The critical temperature values obtained in the present work can be seen in Table \ref{tbl_1}, in comparison with the other methods. We can see from Table \ref{tbl_1} that,  the critical temperatures lie between the variational approach \cite{ref_36} and mean field approximation (MFA) \cite{ref_26}. But we can say that the limiting results of Ising model can be considered as good values in its own class, namely in EFT. The critical value of S-1/2 limit (i.e. $d\rightarrow \infty$) is $t_c=3.025$ is slightly better than the $t_c=3.090$ which is most widely used formulation namely DA \cite{ref_42} within the EFT. Besides, when we compare the  value of the critical temperature in the case $d=0,\Delta=0$, we can see that this value is slightly lower than  the value obtained within the DA \cite{ref_42}, which is $t_c=2.187$. Thus it is expected that, the critical values are slightly lower than the DA \cite{ref_42} for the anisotropic Heisenberg model. Nevertheless we have to say that, the result of the isotropical Heisenberg limit, contradicts to the Mermin-Wagner Theorem \cite{ref_46}, according to which it is not possible to observe any long range order in isotropic Heisenberg model in one and two dimensions. But, this deficiency in limiting case can not prevent the investigation of the effect of the anisotropy in the exchange interaction on the  hysteresis loop behaviors, which is the main topic of this work.

\begin{table}\caption{Critical temperatures in some limiting cases for the square lattice in comparison with corresponding exact, variational and MFA values.}\label{tbl_1}
\center
\begin{tabular}{c|c|c|c|c}
\hline
&Exact&Variational\cite{ref_36}&Present&MFA\cite{ref_26}\\
\hline
$d\rightarrow \infty$&2.269\cite{ref_44}&2.885&3.025&4.0\\
$d=0,\Delta=0$&1.693\cite{ref_45}&2.065&2.149&2.667\\
$d=0,\Delta=1$&0.0\cite{ref_46}&1.492&1.722&2.667
\end{tabular}
\end{table}

\subsection{Hysteresis Behaviors}

In order to investigate the behavior of the hysteresis loops with the variation of the  Hamiltonian parameters, we can use quantities which are related to the shapes of the hysteresis loops, namely hysteresis loop area (HLA), coersive field (CF) and the remanent magnetization (RM).
Since  the evolution of the hysteresis loops with the variation of the temperature is trivial, we inspect the effect of varying crsytal field ($d$) and the anisotropy in the exchange interaction $\Delta$ on the hysteresis loops.

Increasing temperature enhances the thermal fluctuations and this causes to destruct the magnetic order of the system. Hence the HLA, CF and RM values tend to vanish and ferromagnetic hysteresis loops evolve into paramagnetic loops by increasing temperature.

Let us remind the well known effect of the crystal field ($d$) on the order parameter of the system. With given value of $\Delta$, in case of $d>0$,   system will order along the easy axis $z$ up to critical temperature value $t_c$. But for the values that provide $d<0$, the $z$ axis becomes hard direction and when the value of $d$ lowers, the spins tend to align in $xy$ plane, which can be regarded as easy plane. Thus we can say that, the disordered phase of the system can be related to the random alignment of the spins in $z$ direction ($d>0$ and $t>t_c$), as well as random alignment of spins in $xy$ plane ($d<0$).

\begin{figure}[h]\begin{center}
\epsfig{file=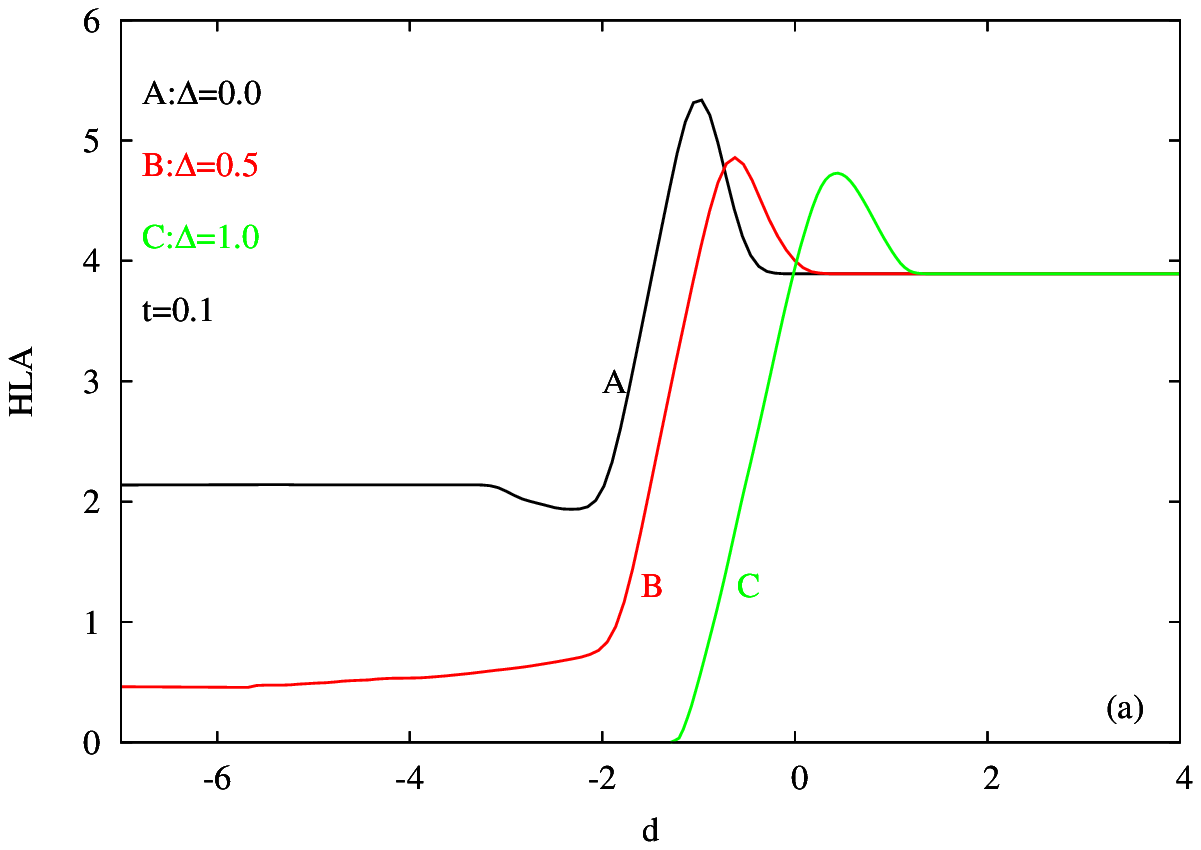, width=6.0cm}
\epsfig{file=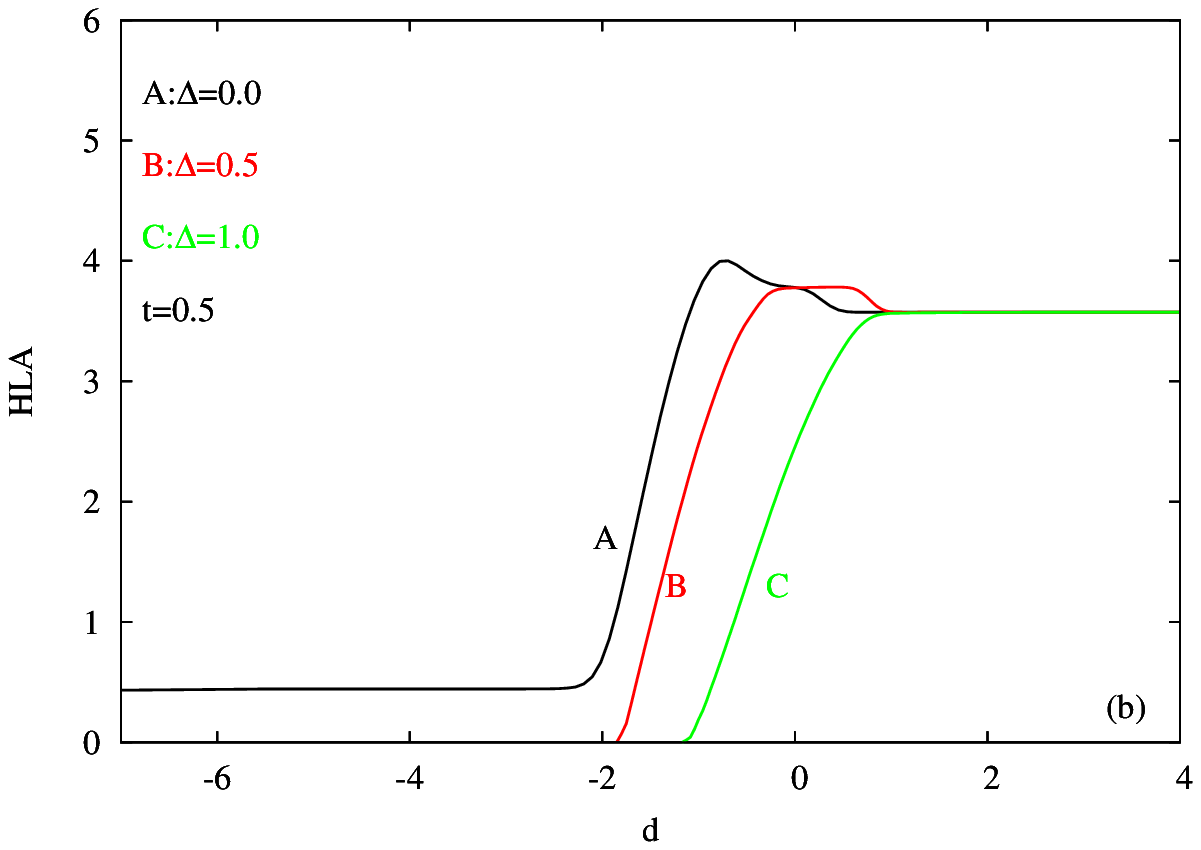, width=6.0cm}
\end{center}
\caption{The variation of the hysteresis loop area with crystal field, for selected values of $\Delta$, at temperature (a) $t=0.1$ and (b) $t=0.5$, for anisotropic S-1 Heisenberg model for square lattice.} \label{sek2}\end{figure}

\begin{figure}[h]\begin{center}
\epsfig{file=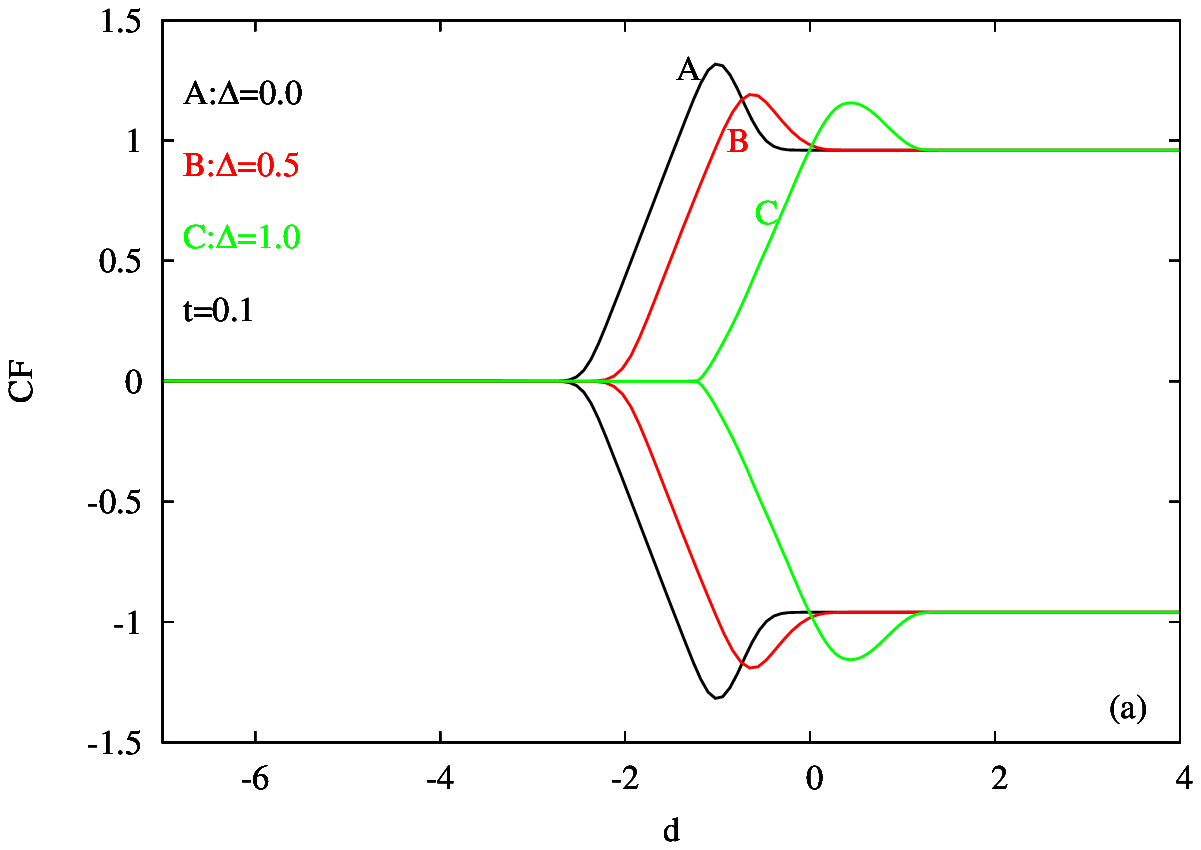, width=6.0cm}
\epsfig{file=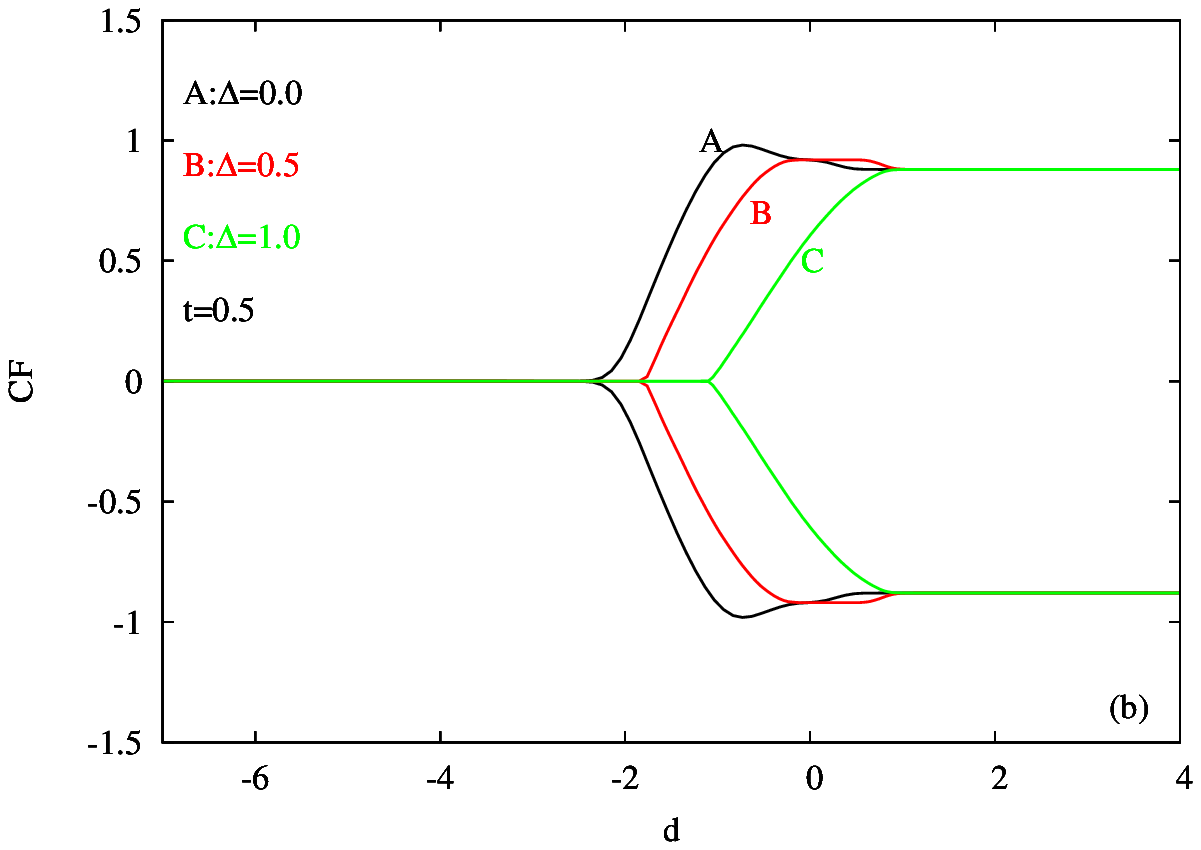, width=6.0cm}
\end{center}
\caption{The variation of the coersive field with crystal field, for selected values of $\Delta$, at temperature (a) $t=0.1$ and (b) $t=0.5$, for anisotropic S-1 Heisenberg model for square lattice.} \label{sek3}\end{figure}

\begin{figure}[h]\begin{center}
\epsfig{file=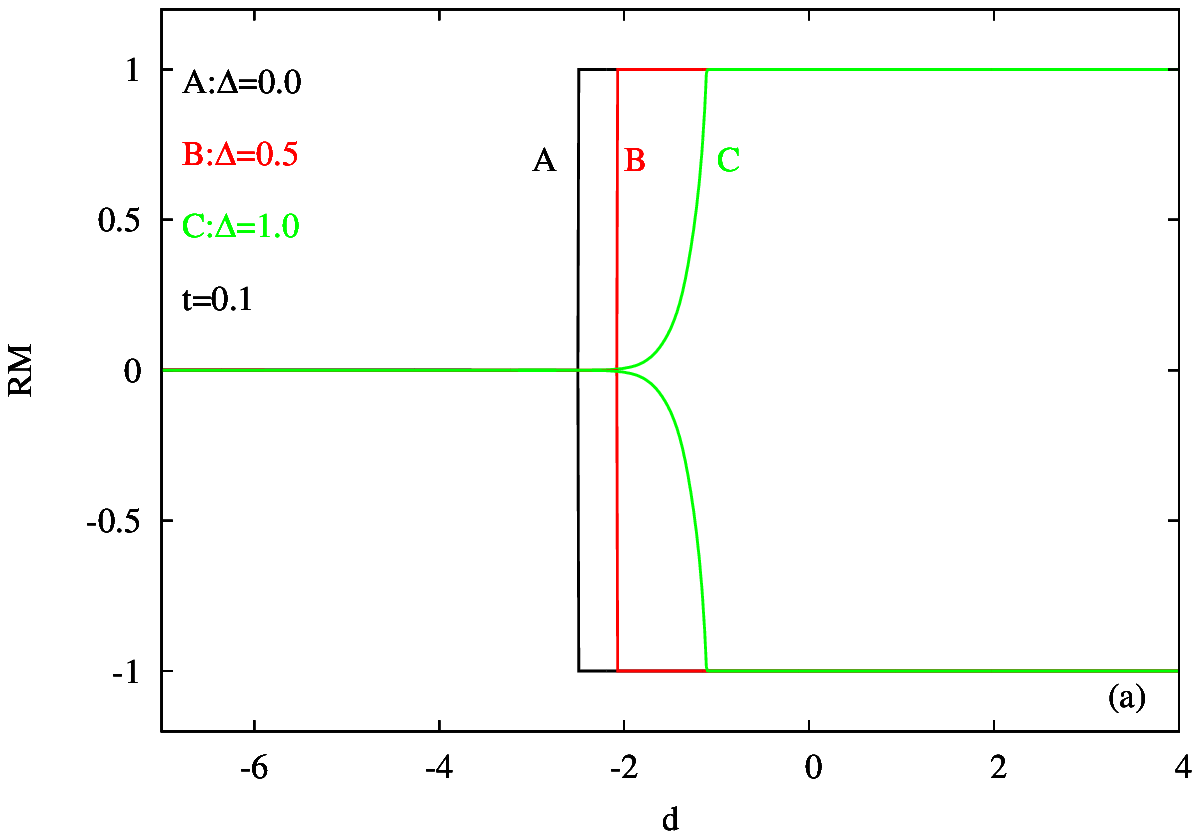, width=6.0cm}
\epsfig{file=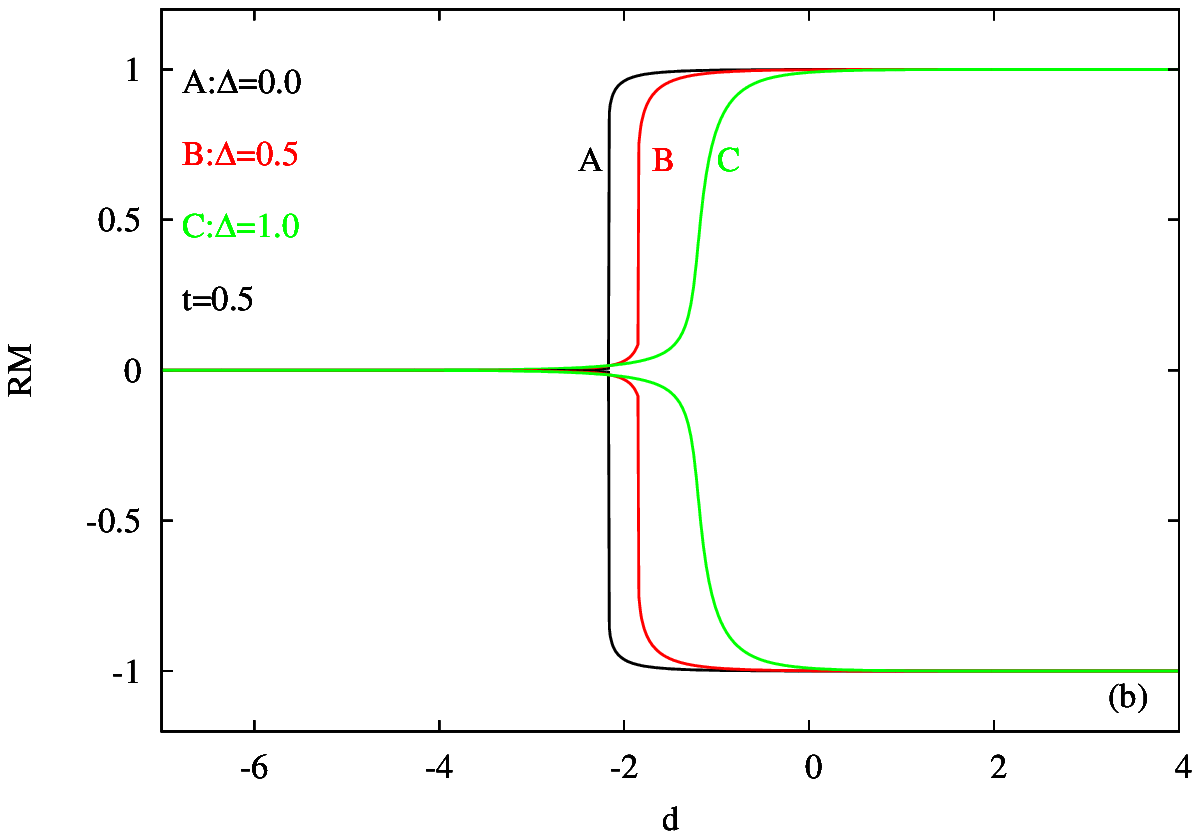, width=6.0cm}
\end{center}
\caption{The variation of the remanent magnetization with crystal field, for selected values of $\Delta$, at temperature (a) $t=0.1$ and (b) $t=0.5$, for anisotropic S-1 Heisenberg model for square lattice.} \label{sek4}\end{figure}

In order to see the effect of the crystal field on the hysteresis behavior, we depict the variation of the HLA with crystal field, for selected values of anisotropy in the exchange interaction in Fig. \ref{sek2}, for selected temperatures as (a) $t=0.1$ and (b) $t=0.5$. The same plots for the CF and RM can be seen in Figs. \ref{sek3} and \ref{sek4}, respectively.

\begin{figure}[h]\begin{center}
\epsfig{file=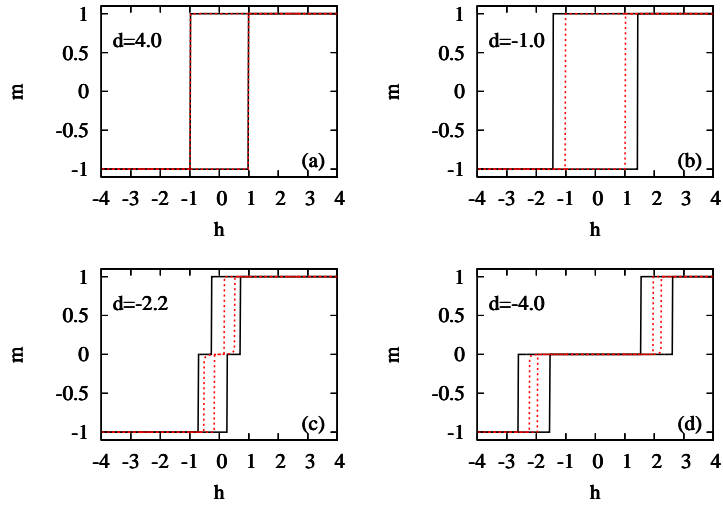, width=10.0cm}
\end{center}
\caption{Hysteresis loops of the XXZ model with $\Delta=0.5$ (shown in dotted red) and Ising model (shown in solid black) on square lattice with temperature  $t=0.1$ for crystal field values (a) $d=4.0$, (b) $d=-1.0$, (c) $d=-2.2$ and  (d) $d=-4.0$.} \label{sek5}\end{figure}

In Fig. \re{sek2}, the curves labeled by A,B,C correspond to the S-1 Ising model,  XXZ model and S-1 isotropic Heisenberg  model, respectively. For positive values of the $d$, varying $\Delta$ has almost no effect on the HLA. At first sight, a qualitative distinction  between the curves A and C stand out in relief. The constant HLA takes place in case of Ising model at large negative values of the crystal field, while this is not the case in isotropic Heisenberg model. By inspecting the corresponding CF and RM curves (see the curves labeled by A in Figs. \re{sek3} (a) and \re{sek4} (a) at large negative values of crystal field) we can say that this nonzero HLA do not originate from the symmetric hysteresis loops around the origin. Because of the large negative values of the crystal field, the system cannot stay in the ferromagnetic phase, then the hysteresis loops of this region correspond to the paramagnetic phase. Since corresponding CF and RM values are zero at large negative values of the crystal field, with nonzero HLA there has to be loops which are not symmetric about the origin in the $m-h$ plane. The hysteresis loops for selected Hamiltonian parameter values from this large negative valued crystal field region can be seen in Fig. \ref{sek5}.

In Fig. \ref{sek5}, we depict the hysteresis loops calculated at the temperature $t=0.1$ for crystal field values (a) $d=4.0$, (b) $d=-1.0$, (c) $d=-2.2$ and  (d) $d=-4.0$ for a square lattice. Each plot contains two representative examples of hysteresis loops corresponding to  $\Delta=0.0$ (Ising model, solid black curve)  and $\Delta=0.5$ (anisotropic Heisenberg model, dotted red curve). First of all, we see from  Fig. \ref{sek5}  that, the loops related to the anisotropic Heisenberg model lie inside of the loops for the Ising model. This is consistent with the results shown in Figs. \ref{sek2}-\ref{sek4}, which conclude that, decreasing anisotropy (i.e. rising $\Delta$) causes to a decline in the HLA, RM and CF. The remarkable result seen in Fig. \ref{sek5} (d) is double hysteresis behavior. When the crystal field decreases, first the hysteresis loops get narrower (compare loops in Figs. \ref{sek5} (a) and (b)  ), then after a specific value of $d$, loops start to split up (see \ref{sek5} (c)), after this double hysteresis loop behavior survives. This result can explain the constant and nonzero behavior of the hysteresis loops with decreasing $d$, at constant temperature.

\begin{figure}[h]\begin{center}
\epsfig{file=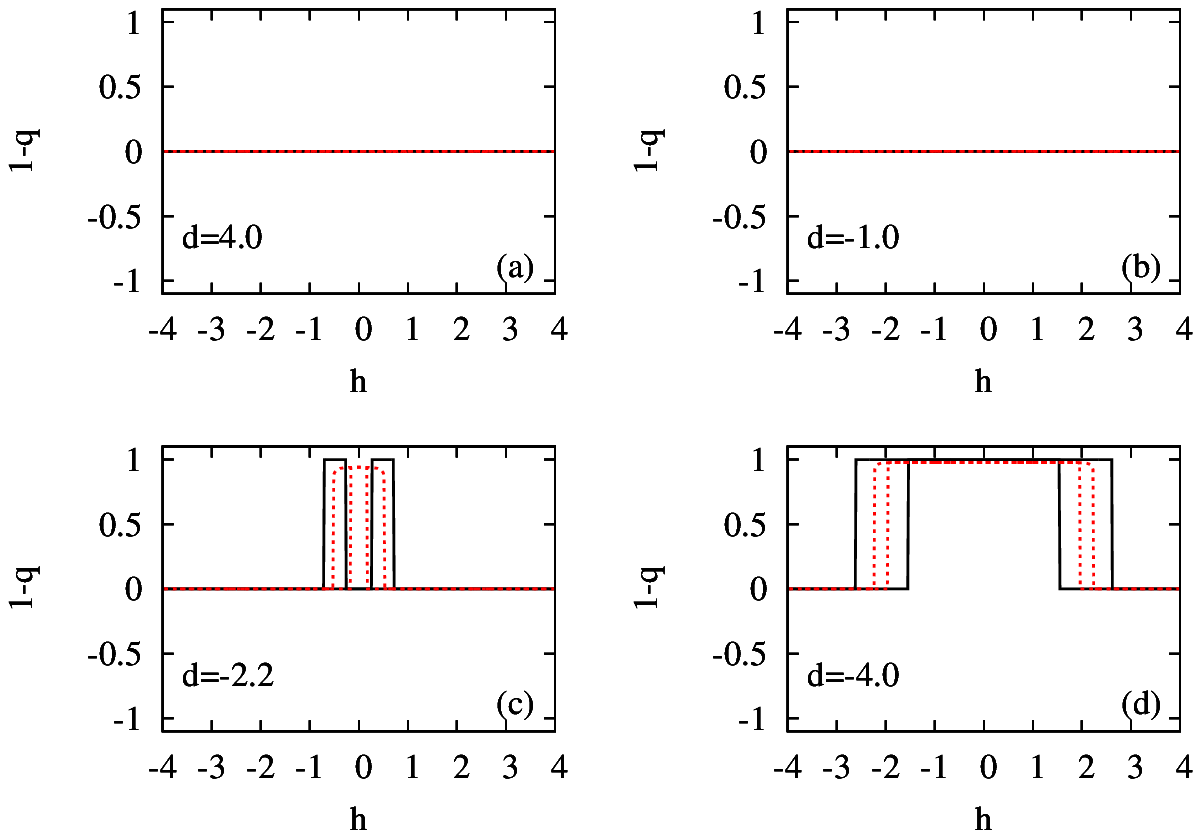, width=10.0cm}
\end{center}
\caption{Hysteresis loops of the XXZ model with $\Delta=0.5$ (shown in dotted red) and Ising model (shown in solid black) on square lattice with temperature  $t=0.1$ for crystal field values (a) $d=4.0$, (b) $d=-1.0$, (c) $d=-2.2$ and  (d) $d=-4.0$.} \label{sek6}\end{figure}

Since the large negative value of $d$ forces the spins to align in $xy$ plane, the double hysteresis loop behavior, which appears in large negative values of the crystal field may come from the forced alignment of the spins from this plane to the $z$ direction. We can see this situation by defining a quantity $(1-q)$, which is the measure of the number of spins aligned in $xy$ plane. In Fig. \ref{sek6} we depict the  "hysteresis" loops for the $1-q$, with same parameters used in Fig. \ref{sek5}. We can see from Figs. \ref{sek6} (a) and (b) that $1-q$ of the system always zero, regardless of the value of the magnetic field. All spins are aligned along parallel to $z$ direction.  But in contrast to this situation, as seen in Fig. \ref{sek6} (d), magnetic field can induce the transition to alignment of the spins from the $xy$ plane to the $z$ axis, when the value of the magnetic field large enough. This situation is consistent with the results given in Fig. \ref{sek5}.   	

We can summarize these results as: the decreasing crystal field gives rise to double hysteresis behavior. This double hysteresis loop behavior comes from the transition between the states which are consist of aligned spins in $xy$ plane and spins in $z$ direction. This mechanism is different from the mechanism which explains the hysteresis behavior in positive valued crystal field.

The double hysteresis loops have been observed experimentally in different systems, for example in Mn-
doped (pb, La) $(Zr,Ti) O_3$ ceramics \cite{ref_47}, and in $Fe_3 O_4 /Mn_3 O_3$ superlattices \cite{ref_48}. In addition, triple hysteresis behaviors have been observed experimentally, such as  single chain magnets with antiferromagnetic interchain
coupling \cite{ref_49} and in molecular-based magnetic materials \cite{ref_50}, $CoFeB/Cu$, $CoNip/Cu$, $FeGa/py$, and
$FeGa/CoFeB$ multilayered nanowires \cite{ref_51}.

In theoretical explanations, this double (or triple) hysteresis loop behaviors mostly attributed to the exchange interaction ratios in the nano material. For instance, the effect of the transverse field  on the hysteresis behavior of the S-1 Ising nanotube has been investigated within the EFT based on a probability distribution method. Double and triple hysteresis loops were obtained \cite{ref_52}. Another work was devoted to the hysteresis behavior of the S-1 Ising ferrielectric cubic nanowire with negative core shell coupling within the  MC simulation. It is observed that, when the absolute value of $J_{shell-core}/J_{core}$ increases, then the hysteresis curve changes from one central loop to triple loops \cite{ref_53}. Also, it was shown within the EFT formulation based on a probability distribution method
that, trimodal random field distribution on the S-1 Ising nanotube could give rise to double hysteresis loops  when the system passes from the ferromagnetic phase to a paramagnetic one \cite{ref_54}. In addition to such studies, we show in the present work that, double hysteresis loop behavior can also occur in bulk systems as explained above. Lastly, one important study deals with the S-1Ising model with transverse crystal field (in $x$ direction) within the EFT with probability distribution technique \cite{ref_55}. In this work the authors cannot find double hysteresis loop behavior due to the transverse alignment of the crystal field.

\begin{figure}[h]\begin{center}
\epsfig{file=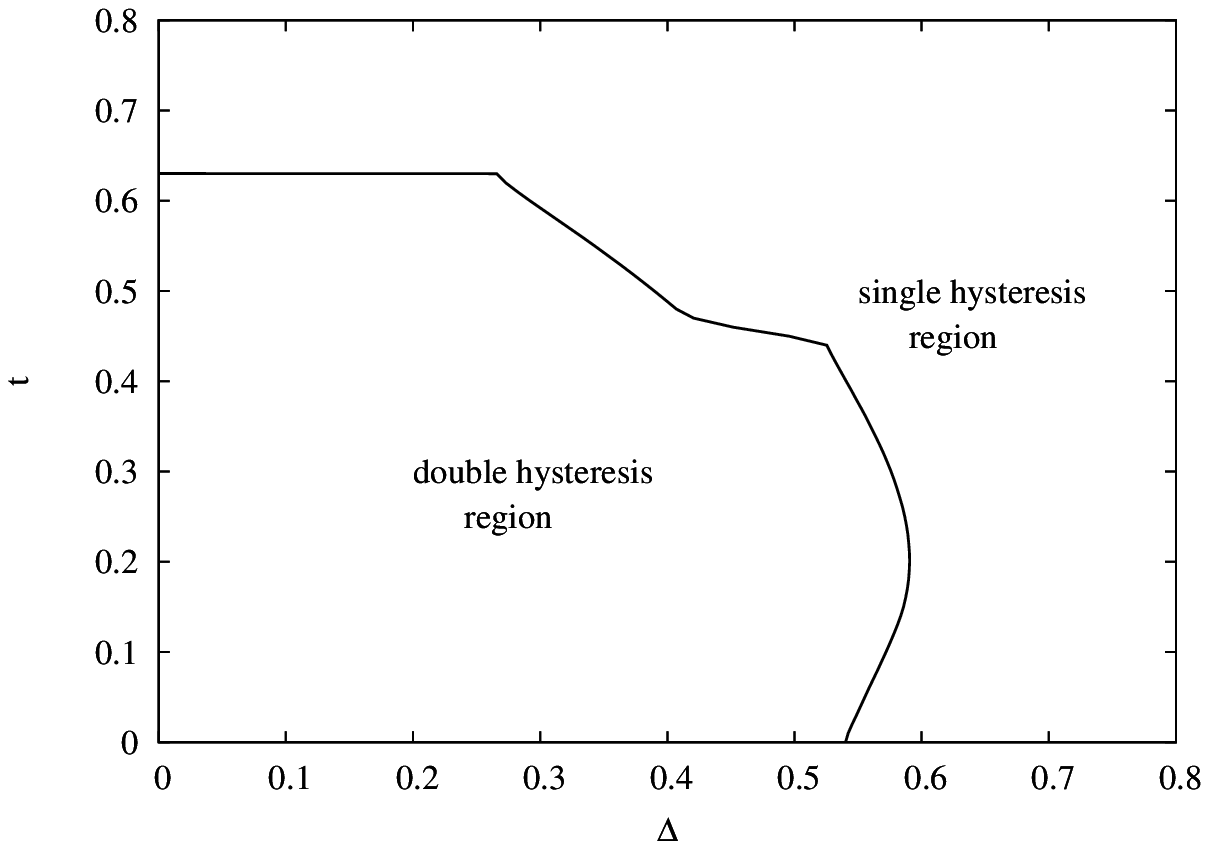, width=10.0cm}
\end{center}
\caption{Double and single hysteresis loop regions of the S-1 anisotropic Heisenberg model on a square lattice in $t-\Delta$ plane. The curve obtained for a large negative value of the crystal field. } \label{sek7}\end{figure}

Lastly, the double hysteresis loop region can be obtained and it is plotted in Fig. \ref{sek7}. In Fig. \ref{sek7}, closed loop seperate the single and double hysteresis loop regions in $(t,\Delta)$ plane. It can be seen from Fig. \ref{sek7} that, double hysteresis loops may occur in low temperature and high anisotropy in the exchange interaction region. After a value of $\Delta=0.591$/$t=0.630$, system cannot show double hysteresis loop behavior at any $t$/$\Delta$. In other words, decreasing anisotropy in the exchange interaction (rising $\Delta$) or rising temperature destroys the double hysteresis loop behavior.

\section{Conclusion}\label{conclusion}

The effects of the crystal field and anisotropy in the exchange interaction of  the anisotropic S-1 Heisenberg model on the hysteresis behavior have been obtained within the effective field theory with two spin cluster. The model has three Hamiltonian parameters as crystal field ($d$), anisotropy in the exchange interaction ($\Delta$) and the temperature ($t$). The model covers the Ising model ($\Delta=0.0$ or $d\rightarrow \infty$) and isotropic Heisenberg model ($\Delta=1.0$) as limiting cases.

After the phase diagram of the system for several values of $\Delta$ given, the effects of  $d$ and $\Delta$ on the hysteresis loops have been  determined. The value of $\Delta$ makes no important difference in the hysteresis loops in low $t$ and positive $d$ region. However in the negative $d$ region one fundamental distinction appears for hysteresis loops that have different values of $\Delta$, namely double hysteresis loop behavior. Double hysteresis loop behavior appears for low values of $t$ and $\Delta$, when $d$ takes large negative values. This behavior is related to the alignment of spins in $xy$ plane and tendency of this alignment to the $z$ direction when external longitudinal magnetic field is applied. It is shown that, this behavior cannot appear in the isotropic Heisenberg model, indeed for the values that provide $\Delta>0.591$. Double hysteresis loop region depicted in ($t,\Delta$) plane has also been investigated in detail and physical explanation given briefly.

We hope that the results  obtained in this work may be beneficial form both theoretical and experimental point of view.
\newpage

\section*{Appendix: Matrix elements of the Hamiltonian}\label{app_a}

Matrix elements of the 2-spin anisotropic Heisenberg Hamiltonian given by Eq. \re{denk2} was defined by Eq. \re{denk5}. The nonzero matrix elements are as follows:

$$
\begin{array}{lcl}
R_{11}&=&-J-(2D+2H+h_1+h_2)\\
R_{22}&=&-(D+H+h_1)\\
R_{24}&=&-2J\Delta\\
R_{33}&=&J-(2D+h_1-h_2) \\
R_{35}&=&-2J\Delta\\
R_{42}&=&-2J\Delta\\
R_{44}&=&-(D+H+h_2)\\
R_{53}&=&-2J\Delta\\
R_{57}&=&-2J\Delta\\
R_{66}&=&-(D-H-h_2)\\
R_{68}&=&-2J\Delta\\
R_{75}&=&-2J\Delta\\
R_{77}&=&J-(2D-h_1+h_2)\\
R_{86}&=&-2J\Delta\\
R_{88}&=&-(D-H-h_1)\\
R_{99}&=&-J-(2D-h_1-h_2)\\
\end{array}
$$

\end{document}